\documentclass[a4paper,11pt]{article}
\usepackage{jheppub} 
\usepackage{lineno}
\usepackage{slashed}

\title{\boldmath Charge-changing weak interactions for right-handed particles in the Standard Model}

\author{J.D. Franson}
\affiliation{University of Maryland Baltimore County,\\
Baltimore, Maryland USA }

\emailAdd{jfranson@umbc.edu}

\abstract{Experiments have shown that the charge-changing weak interaction is purely left-handed, which is taken into account in the Standard Model by the inclusion of a left-handed projection operator in the Lagrangian.  Nevertheless, it will be shown here that the Standard Model predicts charge-changing weak interactions for right-handed fermions that can be larger than those for left-handed fermions if the mass is sufficiently large, as is the case for the top quark.  Here we are using the conventional terminology in which a massive fermion with its spin parallel to its momentum is referred to as being right-handed in the relativistic limit, where it is in an approximate eigenstate of the chirality operator.  These effects are due to the way in which the field of the W boson is quantized, which gives a divergent tensor product in the Feynman propagator in the unitary gauge.  It will be shown that the off-diagonal terms in the propagator can convert a left-handed projection operator into a right-handed projection operator, which allows an interaction with right-handed fermions even though the Lagrangian is left-handed.   Experiments to date have only demonstrated charge-changing weak interactions for left-handed particles, and an alternative quantization approach that eliminates the  divergent off-diagonal terms in the W boson propagator and avoids these difficulties will be considered.  The alternative approach appears to be in agreement with existing experiments, but  additional high-energy experiments may be required in order to distinguish its predictions from those of the Standard Model. }

\begin{document}
\maketitle
\flushbottom

\section{Introduction}
\label{sec:intro}

The pioneering experiment of Wu and her colleagues \cite{wu1957} showed that the weak interaction responsible for beta decay is purely left-handed, which violates parity \cite{lee1956}.  This is taken into account in the Standard Model by the $SU{(2)_L}$  symmetry of the Lagrangian, where an interaction term that is proportional to 
$(1 - {\gamma ^5})/2$    projects out the left-handed component of a fermion \cite{sudarshan1958,glashow1959,salam1959,weinberg1967,maiani2016,schwartz2014,peskin1995,griffiths2008,weinberg1996,nagashima2013,bardin1999}.  ${(\gamma ^5}$ is one of the Dirac matrices.)  As a result, it seems reasonable to conclude that “charge-changing weak interactions couple only to left-handed-polarized fermions” \cite{peskin2017}.  

\begin{figure}[tbp]
\centering
\includegraphics[width=0.35\textwidth]{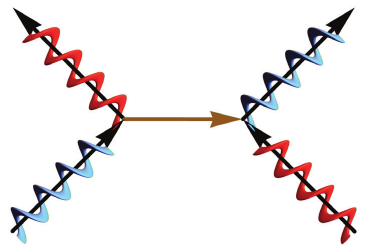}
\qquad
\caption{Scattering of a right-handed top quark (blue) and a left-handed bottom quark (red), with the exchange of a virtual W boson (brown).  The Standard Model predicts a cross section for this process that is much larger than that for a left-handed top quark, as will be shown below.}
\label{figure 1}
\end{figure}

Nevertheless, it will be shown here that the Standard Model predicts charge-changing weak interactions for right-handed fermions that can be larger than those for left-handed fermions if the mass is sufficiently large, as is the case for the top quark in figure \ref{figure 1}.  Here we are using the conventional terminology in which a massive fermion with its spin parallel to its momentum is referred to as being right-handed in the relativistic limit, where it is in an approximate eigenstate of the chirality operator $ \gamma ^5 $.

The charge-changing weak interaction is due to the exchange of virtual W  bosons, and the canonical quantization of the field in the unitary gauge gives a tensor product
${q_\mu }{q_\nu }/{M_W}^2$  in the Feynman propagator for the W  boson \cite{peskin1995,griffiths2008,weinberg1996,nagashima2013,bardin1999}.  (Here ${q_\mu }$ is the 4-vector momentum transfer and 
${M_W}$  is the mass of the W  boson.)  It will be shown that the off-diagonal terms in the propagator can convert the left-handed projection operator 
$(1 - {\gamma ^5})/2$   into the right-handed projection operator 
$(1 + {\gamma ^5})/2$.  This allows an interaction with right-handed fermions even though the Lagrangian is left-handed.  These effects are unrelated to proposals for right-handed interactions that go beyond the Standard Model 
\cite{wilczek1975,derujula1977,arguelles2016}.

These results are possible because the spinor for a massive “right-handed” fermion will always contain a small left-handed component even at relativistic velocities.  The tensor product 
${q_\mu }{q_\nu }/{M_W}^2$  diverges in the limit of high energies, and that term can multiply the vanishingly-small left-handed component to give a large but finite contribution to the scattering amplitude for a right-handed fermion at high energies. This is not merely a semantic issue, however, since experiments to date have shown charge-changing weak interactions only for left-handed fermions \cite{wu1957,quin1989,ashery2017}, and it is an open experimental question as to whether or not right-handed fermions can have large charge-changing cross sections at sufficiently high energies.

The presence of the 
${q_\mu }{q_\nu }/{M_W}^2$  term in the conventional Feynman propagator is due to the fact that only 3 of the 4 components of the field of a massive vector boson are quantized in the unitary gauge of the Standard Model \cite{weinberg1995}. An alternative approach that quantizes all 4 components of the field in a covariant way using the indefinite metric \cite{gupta1950,gupta1957,bleuler1950,gupta1977,cohen-tannoudji1989,itzykson1980} will be considered.  This eliminates the divergent tensor product in the propagator and ensures that the charge-changing weak interaction is negligible for right-handed fermions.

The origin of these effects can be further understood from the lack of rotational symmetry of the internal degrees of freedom of the field of the W  boson after canonical quantization, as will be discussed in more detail below.  This suggests that the spin of a W boson may not be conserved, which provides an intuitive explanation for how its propagator can convert a left-handed projection operator into a right-handed projection operator.  All of these effects depend on the fact that the weak current ${j^\mu }$  is not conserved 
\cite{paschos2007}, since the Lagrangian for the Proca equation by itself is not gauge invariant.  The same issues occur if the propagator is derived using Feynman path integrals \cite{weinberg1995,feynman1948,feynman1950,feynman1965,srednicki2007}, although their origin is not as apparent in that case.  

The alternative quantization approach maintains the rotational symmetry of the system, which avoids these difficulties, and it appears to be in agreement with existing experiments.  Additional high-energy experiments may be needed in order to distinguish between the predictions of the alternative approach and those of the Standard Model.

The unitary gauge will be used throughout this paper, where the gauge parameter 
$\xi  \to \infty $.  This has the advantage that it eliminates all nonphysical particles such as the Goldstone boson \cite{bardin1999}.  The same results can be obtained using the ‘t Hooft-Feynman gauge  
$ ( \xi =1) $ \cite{thooft1971,fujikawa1972}, where the vertex factor for the nonphysical Goldstone boson explicitly includes a right-handed term that is proportional to $(1 + {\gamma ^5})/2$  \cite{bardin1999,nagashima2013}.  It should be emphasized, however, that there are no nonphysical particles involved in the analysis here, which is based on the unitary gauge.

The remainder of the paper begins with a brief review of the canonical quantization of a massive vector field such as the W  or the 
$\text{Z}^0$  in the unitary gauge.  The tensor product in the Feynman propagator will then be used to rewrite the conventional scattering amplitude in an equivalent form that includes both right-handed and left-handed projection operators.  The lack of rotational symmetry of the internal degrees of freedom of the field after canonical quantization will be discussed.  An alternative quantization approach will then be considered in which all four components of the field are quantized, which maintains the rotational symmetry of the system and gives charge-changing weak interactions only for left-handed fermions. Finally, the feasibility of experimentally distinguishing between the predictions of the Standard Model and those of the alternative quantization approach will be discussed.

\section{Canonical quantization}

The effects of interest in this paper are due to the way in which the field of a massive vector boson is quantized in the unitary gauge of the Standard Model, which will be briefly reviewed in this section.  The same propagator can be obtained using canonical quantization \cite{weinberg1995} or by using Feynman path integrals \cite{weinberg1995,feynman1948,feynman1950,feynman1965,srednicki2007}.  The focus in this section will be on the canonical quantization procedure, where the origin of the tensor product in the propagator is most apparent.

The discussion at this point will closely follow that in Weinberg’s text \cite{weinberg1995}.  For simplicity, a real vector field $V^\mu$ with mass $M$  will be considered, which can be described in the unitary gauge by a Lagrangian density 
$\mathcal{L}$  given by  
\begin{equation}     
\mathcal{L} =  - \frac{1}{4}{F_{\mu \nu }}{F^{\mu \nu }} + \frac{1}{2}{M^2}{V_\mu }{V^\mu } - {j_\mu }{V^\mu }.
\label{eqL}
\end{equation}
Similar results can be obtained for a complex field.  Here $M$   is the mass of the boson,\\ 
${F_{\mu \nu }} = {\partial _\mu }{V_\nu } - {\partial _\nu }{V_\mu }$  as in electromagnetism, and  
${j^\mu }$ corresponds to the weak current associated with another particle such as a fermion.  The metric tensor ${\eta ^{\mu \nu }}$  will be chosen to have diagonal elements of $(1, - 1, - 1, - 1)$  and units with $c = \hbar  = 1$  will be used.

The time derivative of $V^0$   cancels out of the Lagrangian, and the corresponding Euler-Lagrange equation gives the constraint
\begin{equation}
{V^0} = \frac{1}{{{M^2}}}\left( {\nabla  \cdot \boldsymbol{\Pi } + j^0} \right),
\label{eqV0}
\end{equation}
where $\boldsymbol{\Pi}$ is the conjugate field.  As usual, bold letters represent 3-vectors whose indices will be labeled by Roman letters.

Eq.\eqref{eqV0} shows that $V^0$  does not correspond to an independent degree of freedom, and only the three spatial components 
${V^i}$  are quantized in the canonical approach.  The other Euler-Lagrange equations can be used to show that
\begin{equation}
{\partial _\mu }{V^\mu } = 0,
\label{eqlorentz}
\end{equation}
provided that ${\partial _\mu }{j^\mu } = 0$ \cite{weinberg1995}, which corresponds to a generalized Lorentz condition.  It is assumed that the constraint of equations \eqref{eqV0} and \eqref{eqlorentz} reduces the number of independent polarization vectors from 4 to 3.

In the absence of interactions $({j^\mu } = 0),$  the field operator that is consistent with equations \eqref{eqL} through \eqref{eqlorentz} has the form \cite{weinberg1995}
\begin{equation}
{V^\mu }(x) = \frac{1}{{{{\left( {2\pi } \right)}^3}}}\sum\limits_{s = 1}^3 {\int {\left( {\frac{{{d^3}\boldsymbol{p}}}{{\sqrt {2{p^0}} }}} \right)} } {\lambda ^\mu }(\boldsymbol{p},s)a(\boldsymbol{p},s){e^{ - ip \cdot x}} + h.c. 
\label{eqVmu}
\end{equation}
The three polarization vectors $ \lambda ^ \mu (\boldsymbol{p},s) $ can be defined in the rest frame of the particle as
\begin{equation}
{\lambda ^\mu }(0,1) = \left( {\begin{array}{*{20}{c}}
0\\
1\\
0\\
0
\end{array}} \right)\quad {\lambda ^\mu }(0,2) = \left( {\begin{array}{*{20}{c}}
0\\
0\\
1\\
0
\end{array}} \right)\quad {\lambda ^\mu }(0,3) = \left( {\begin{array}{*{20}{c}}
0\\
0\\
0\\
1
\end{array}} \right).
\label{eqlambdamu}
\end{equation}
A subsequent boost to a coordinate frame where the particle has momentum $\boldsymbol{p}$ is given by a Lorentz transformation $ \Lambda $ with 
\begin{equation}
\lambda ^\mu (\boldsymbol{p},s) = {\Lambda ^\mu} _\nu {\lambda ^\nu }(0,s).
\label{eqlambdamup}
\end{equation}
The annihilation operators $a(\boldsymbol{p},s)$  are assumed to obey the usual commutation relations given by
\begin{equation}
[a(\boldsymbol{p},s),{a^\dag }(\boldsymbol{p}',s')] = {(2\pi )^3}{\delta ^{(3)}}(\boldsymbol{p} - \boldsymbol{p}'){\delta _{ss'}}.
\label{eqcommutators}
\end{equation}

Equations \eqref{eqVmu} through \eqref{eqcommutators}  can be used to show that the Feynman propagator $\Delta ^U _{\mu \nu }$  for the field is given by \cite{peskin1995,griffiths2008,nagashima2013,weinberg1995}
\begin{equation}
\Delta _{\mu \nu }^U =  - i\frac{{\left( {{\eta _{\mu \nu }} - {q_\mu }{q_\nu }/{M^2}} \right)}}{{\left( {{q^2} - {M^2} + i\varepsilon } \right)}}.
\label{eqpropu}
\end{equation}
Here the superscript $U$  indicates that this is the Feynman propagator in the unitary gauge of the Standard Model.  The presence of the tensor product  
${q_\mu }{q_\nu }/{M^2}$  is due to the fact that 
${V^0}$  was not quantized, and it causes a number of technical difficulties since it diverges as  
${q_\nu } \to \infty .$ The off-diagonal terms in the propagator will also be found to be responsible for the interaction with right-handed fermions in the analysis that follows.

Stueckelberg \cite{stueckelberg1938,ruegg2004} considered a massless vector field and then included an interaction with an additional scalar field, which gave the particle a mass similar to the Higg's mechanism.  The constraint in eq.\ \eqref{eqV0} could also be taken into account by adding a Lagrange multiplier or a gauge-fixing term to the Lagrangian \cite{dirac1950,faddeev1967,senjanovic1976,grosse-knetter1993}.  All of these procedures give the same Feynman propagator and are therefore physically equivalent to the use of canonical quantization.

The unitary gauge has the advantage that the nonphysical Goldstone boson has been absorbed into the longitudinal component of 
${V^\mu }$  and it does not contribute to any Feynman diagrams.  As a result, the propagator of eq.\ \eqref{eqpropu} is what usually appears in the Feynman rules for the Standard Model \cite{peskin1995,griffiths2008,weinberg1996,nagashima2013,bardin1999}.

It should be noted that the Lorentz condition 
${\partial _\mu }{V^\mu } = 0$  in eq.\ \eqref{eqlorentz} was derived under the assumption that the weak current is conserved with ${\partial _\mu }{j^\mu } = 0$  \cite{weinberg1995,greiner1996}. The implications of this will be discussed in section 4.3.

\section{Scattering amplitudes for right-handed fermions}

It will be shown in this section that right-handed fermions can have larger charge-changing weak interactions than left-handed fermions if their mass is sufficiently large.  This can be understood from the fact that the Feynman propagator for the W boson can convert a left-handed projection operator into a right-handed projection operator, as will also be shown.

\subsection{Equivalent form of the scattering amplitude}

The conventional expression for the lowest-order scattering amplitude involving the exchange of a virtual W  boson will now be rewritten in an equivalent form using the  ${q_\mu }{q_\nu }/{M^2}$  term in the Feynman propagator.  The lowest-order Feynman diagram for the scattering of an electron and an electron neutrino with the exchange of a virtual W  boson is shown in figure 2(a), while the analogous process involving the scattering of a top and bottom quark is illustrated in figure 2(b).  The bottom quark has a charge of -1/3 while the top quark has a charge of +2/3, so that the exchange of a $\text{W}^+$  boson maintains charge conservation.  The effects of interest will be found to be strongly dependent on the mass of the fermions, and the analysis will focus primarily on the process shown in figure 2(b) due to the large mass of the top quark.

\begin{figure}[tbp]
\centering
\includegraphics[width=0.63\textwidth]{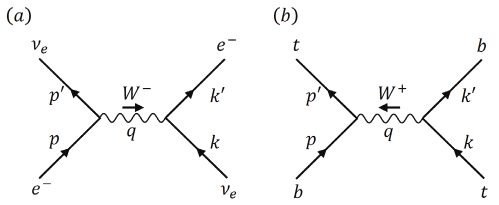}
\qquad
\caption{(a)  Scattering of an electron ${e^ - }$  and an electron neutrino ${\nu _e}$  with the exchange of a virtual $\text{W}^-$  boson.  (b)  Scattering of a bottom quark b with a top quark t.  The 4-momenta of the particles are labeled 
$p,\;p',\;k,\;{\rm{and}}\;k',$  while $q$  is the 4-momentum transfer. Time flows in the upward direction.}
\label{figure 2}
\end{figure}

The vertex factor ${V_W}$  associated with the exchange of a W  boson is given by \cite{griffiths2008,nagashima2013}
\begin{equation}    
{V_W} =  - i\frac{{{g_W}}}{{2\sqrt 2 }}{\gamma ^\mu }\left( {1 - {\gamma ^5}} \right),
\label{eqVw}
\end{equation}
where ${g_W}$  is the electroweak coupling constant.  Combining the vertex factor ${V_W}$  with the propagator of eq.\ \eqref{eqpropu} gives a lowest-order conventional scattering amplitude 
$\mathcal{M}$  of the form
\begin{equation}
\mathcal{M} = {\mathcal{M}}_1 + {\mathcal{M}}_2,
\label{eqM}
\end{equation} 
where
\begin{eqnarray}
{\mathcal{M}}_1 & = &  - ig{'^2}[{{\bar u}_t}(p'){\gamma ^\mu }(1 - {\gamma ^5}){u_b}(p)]\left( {\frac{{{\eta _{\mu \nu }}}}{{{q^2} - {M_W}^2}}} \right)  [{{\bar u}_b}(k'){\gamma ^\nu }(1 - {\gamma ^5}){u_t}(k)]
  \nonumber ,\\[2pt]
{\mathcal{M}}_2 & = & ig{'^2}[{{\bar u}_t}(p'){\gamma ^\mu }(1 - {\gamma ^5}){u_b}(p)]\left( {\frac{{{q_\mu }{q_\nu }/{M_W}^2}}{{{q^2} - {M_W}^2}}} \right)
 [{{\bar u}_b}(k'){\gamma ^\nu }(1 - {\gamma ^5}){u_t}(k)].
\label{eqM12}
\end{eqnarray}
Here 
$g' \equiv  - i{g_W}/2\sqrt 2 $  while 
${u_t}(k)$  denotes a Dirac spinor corresponding to a top quark with 4-momentum  $k,$ with a similar notation for the other spinors. 
${M_W}$  is the mass of a  W  boson while
$q = p' - p$  is the 4-momentum transfer.

${\mathcal{M}_1}$ involves only left-handed currents, while the presence of the tensor product \\ 
${q_\mu }{q_\nu }/{M_W}^2$  in 
${\mathcal{M}_2}$  makes the situation more subtle.  In order to see this, ${\mathcal{M}_2}$  will be put into an equivalent form using
\begin{equation}
 \{ {\gamma ^\nu },{\gamma ^5}\}  = 0,
 \label{eqcom2}
\end{equation}
from which it follows that
\begin{equation}    
{\gamma ^\nu }(1 - {\gamma ^5}) = (1 + {\gamma ^5}){\gamma ^\nu }.
\label{eqgam5}
\end{equation}

The momentum transfer ${q_\nu }$  is contracted with ${\gamma ^\nu }$  in eq.\  \eqref{eqM12}, so that the right-hand side of the equation for ${\mathcal{M}_2}$  contains a product of factors that will be denoted by ${F_R :}$ 
\begin{eqnarray}  
{F_R} & \equiv & {q_\nu }[{{\bar u}_b}(k'){\gamma ^\nu }(1 - {\gamma ^5}){u_t}(k)] \nonumber \\[2pt]
 & = & {{\bar u}_b}({k'})\slashed{q}(1 - {\gamma ^5}){u_t}(k) \nonumber \\[2pt]
&  = & {{\bar u}_b}(k')(\slashed{k}-\slashed{k}')(1 - {\gamma ^5}){u_t}(k).
\label{eqFR}
\end{eqnarray} 
Here the Feynman slash notation
$\slashed{q} \equiv {\gamma ^\nu }{q_\nu } $  has been used along with  
$q = k - k'.$  Eq.\ \eqref{eqgam5} can be used to rewrite ${F_R}$  in the form 
\begin{equation}
{F_R} =  - {\bar u_b}(k')\slashed{k}'(1 - {\gamma ^5}){u_t}(k) + {\bar u_b}(k')(1 + {\gamma ^5})\slashed{k}{u_t}(k).
\label{eqFR2}
\end{equation}

Since the spinors are positive-energy solutions to the Dirac equation, they satisfy the conditions
\begin{eqnarray}    
\slashed{k}{u_t}(k) & = & {m_t}{u_t}(k), \nonumber \\[2pt]
{{\bar u}_b}(k')\slashed{k}' & = & {m_b}{{\bar u}_b}(k'),
\label{eqkslash}
\end{eqnarray}
where ${m_t}$  and ${m_b}$  are the masses of the top and bottom quarks.  Inserting eq.\  \eqref{eqkslash} into eq.\ \eqref{eqFR2} gives
\begin{equation}
{F_R} = {\bar u_b}(k')[ - {m_b}(1 - {\gamma ^5}) + {m_t}(1 + {\gamma ^5})]{u_t}(k).\label{eqFR3}
\end{equation}
The product of factors in the left-hand side of 
${\mathcal{M}_2}$  can be rewritten in the same way, and combining these results gives
\begin{eqnarray}
{\mathcal{M}_2} & = & ig{'^2}{{\bar u}_t}(p')\left[ {\frac{{{m_b}}}{{{M_W}}}(1 + {\gamma ^5}) - \frac{{{m_t}}}{{{M_W}}}(1 - {\gamma ^5})} \right]{u_b}(p) \nonumber \\[2pt]
 & \times & \left( {\frac{1}{{{q^2} - {M_W}^2}}} \right){{\bar u}_b}(k')\left[ {\frac{{{m_b}}}{{{M_W}}}(1 - {\gamma ^5}) - \frac{{{m_t}}}{{{M_W}}}(1 + {\gamma ^5})} \right]{u_t}(k).
\label{eqM2}
\end{eqnarray}
Similar techniques were used in ref.\ \cite{peskin1995} to show that the ${R_\xi }$  gauges are equivalent for a simplified model.

The mass of the top quark is much larger than that of the bottom quark, so that eq.\ \eqref{eqM2} is given to a good approximation by
\begin{eqnarray}
{\mathcal{M}_2}  \approx  ig{'^2}{\left( {\frac{{{m_t}}}{{{M_W}}}} \right)^2}\left[ {{{\bar u}_t}(p')(1 - {\gamma ^5}){u_b}(p)} \right] 
    \left( {\frac{1}{{{q^2} - {M_W}^2}}} \right)\left[ {{{\bar u}_b}(k')(1 + {\gamma ^5}){u_t}(k)} \right].
\label{eqM2a}
\end{eqnarray}
Similar results can be obtained for antiparticles using  
$\not{k}v(k) =  - mv(k),$ where  
$v(k)$ is the spinor for the antiparticle.

It can be seen that equations \eqref{eqM2} and \eqref{eqM2a} involve both right-handed and left-handed interactions proportional to 
$(1 \pm {\gamma ^5})/2.$  Both of the factors in square brackets in eq.\ \eqref{eqM2a} project out the right-handed component of the top quark, since the first factor can be rewritten using
\begin{eqnarray}
{{\bar u}_t}(p')(1 - {\gamma ^5}) & = & {u^\dag }_t(p'){\gamma ^0}(1 - {\gamma ^5}) \nonumber \\[2pt] & = & {u^\dag }_t(p')(1 + {\gamma ^5}){\gamma ^0} \nonumber \\[2pt]
 & = & {\left[ {(1 + {\gamma ^5}){u_t}(p')} \right]^\dag }{\gamma ^0}.
\label{equt}
\end{eqnarray}
In the same way, both sides of the equation also project out the left-handed component of the bottom quark.

The cross section from $\mathcal{M}_2$  is proportional to 
${({m_t}/{M_W})^4} \approx 21.3,$  which shows that the right-handed scattering process shown in figure \ref{figure 1} is an order of magnitude larger than the  left-handed scattering from ${M_1},$ where the corresponding factor is 1.  The feasibility of experimentally observing these effects will be discussed in section 5.2.

Although the unitary gauge is used throughout this paper, it may be useful to consider the situation in the ‘t Hooft-Feynman gauge, where the Goldstone boson is not absorbed into the longitudinal component of the W  boson.  This gives rise to an additional Feynman diagram similar to figure 2(b) with the  $\text{W}^+$  boson replaced with a Goldstone boson ${\phi ^ + }.$ The vertex factor ${V_G}$  for the charged Goldstone boson is given by \cite{bardin1999,nagashima2013}
\begin{equation}
{V_G} =  - i\frac{{{g_W}}}{{2\sqrt 2 }}\left( {\frac{{{m_b}}}{{{M_W}}}(1 - {\gamma ^5}) - \frac{{{m_t}}}{{{M_W}}}(1 + {\gamma ^5})} \right).
\label{eqVg}
\end{equation}
It can be seen that ${V_G}$  contains interactions with both left-handed and right-handed fermions in agreement with the results obtained here using the unitary gauge in eq.\ \eqref{eqM2}, as would be expected from gauge invariance. 

The Goldstone boson is nonphysical because its mass depends on the gauge parameter  $\xi.$  One might argue that the appearance of the right-handed projection operator in the \\
‘t Hooft-Feynman gauge must be an unobservable artifact due to the nonphysical nature of the Goldstone boson or the subtleties of gauge transformations.  That argument may seem all the more plausible given that the vertex factor ${V_W}$  in the unitary gauge depends only on  $(1 - {\gamma ^5})/2.$  

These conceptual difficulties have been avoided here  by using the unitary gauge throughout, where there  are no nonphysical particles.  The Feynman diagram of figure \ref{figure 1} is just as observable as any other Feynman diagram in the unitary gauge.

\subsection{Projection operators}

The results of the previous section show that the scattering amplitude includes both left-handed and right-handed projection operators, even though the Lagrangian is purely left-handed.  Additional insight into this situation can be obtained by considering the effects of the Feynman propagator on the projection operators  $(1 \pm {\gamma ^5})/2.$

First consider the effects of projection operators  
${P_L}$ and ${P_L}'$  defined by
\begin{eqnarray}
{P_L} & \equiv & \frac{{(1 - {\gamma ^5})}}{2}, \nonumber \\[2pt]
{P_L}' & \equiv & \frac{{\slashed{k}}}{m}\frac{{(1 - {\gamma ^5})}}{2}.
\label{eqPL}
\end{eqnarray}
${P_L}$ is the usual left-handed projection operator while  ${P_L}'$ is a modified projection operator obtained by acting on the left with 
$\slashed{k}/m$  as in the scattering amplitude of eq.\ \eqref{eqFR}.  The operators  ${P_R}$ and ${P_R}'$  can be defined in a similar way using 
$(1 + {\gamma ^5})/2.$ 

The effects of the operator ${P_L}'$  acting on a spinor $u(k)$  that is a positive-energy solution to the Dirac equation can be seen from
\begin{eqnarray}
{P_L}'u(k) & = & \frac{{\slashed{k}}}{m}\frac{{(1 - {\gamma ^5})}}{2}u(k) \nonumber\\[2pt]
& = & \frac{{(1 + {\gamma ^5})}}{2}\frac{{\slashed{k}}}{m}u(k) \nonumber \\[2pt]
 &  = & \frac{{(1 + {\gamma ^5})}}{2}u(k)  \nonumber\\[2pt]
 & = & {P_R}u(k).
\label{eqPLprime}
\end{eqnarray}
Here  $m$ is the mass of a fermion with momentum 
$k$  and eq.\ \eqref{eqgam5} has been used.  A similar result can be shown for ${P_R}'.$  

Since $u(k)$  is the most general positive-energy solution to the Dirac equation, these results can be summarized by operator relations of the form
\begin{eqnarray}
\frac{{\slashed{k}}}{m}{P_L} & = & {P_R}, \nonumber \\[2pt]
\frac{{\slashed{k}}}{m}{P_R} & = & {P_L},
\label{eqconvert}
\end{eqnarray}
where it is understood that the projection operators act on positive-energy spinors with momentum  $k.$ Eq.\ \eqref{eqconvert} shows that the tensor product in the Feynman propagator has the effect of converting a left-handed projection operator into a right-handed projection operator and vice versa.

\begin{figure}[tbp]
\centering
\includegraphics[width=0.68\textwidth]{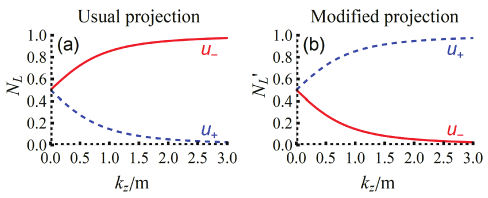}
\qquad
\caption{ (a) Effects of the usual left-handed projection operator ${P_L} \equiv (1 - {\gamma ^5})/2.$  The normalized projection 
${N_L} \equiv |{P_L}{u_ \pm }{|^2}/|{u_ \pm }{|^2}$  is plotted as a function of ${k_z}/m,$  where the red (solid) line corresponds to a negative-helicity spinor ${u_ - }$  while the blue (dashed) line corresponds to a positive-helicity spinor ${u_ + }.$  (b) Effects of the modified projection operator 
${P_L}' \equiv \slashed{k} (1 - {\gamma ^5})/2m,$  where the normalized projection  
${N_L}' \equiv |{P_L}'{u_ \pm }{|^2}/|{u_ \pm }{|^2}$ is plotted as a function of ${k_z}/m.$   It can be seen that  the operator $\not{k}/m$  from the Feynman propagator converts a left-handed projection operator into a right-handed projection operator, in agreement with eq.\ \eqref{eqconvert}.  (dimensionless units.) }
\label{figure 3}
\end{figure}

These results are illustrated in figure \ref{figure 3} for spinors ${u_ + }({k_z})$ and ${u_ - }({k_z})$  with positive and negative helicities, respectively.  For simplicity, the fermion will be assumed to be moving along the positive z axis, in which case \cite{baym1973,bjorken1964}
\begin{eqnarray}   
{u_ + }({k_z}) & = & \sqrt {\frac{{E + m}}{2}} \left( {\begin{array}{*{20}{c}}
1\\
0\\
{{k_z}/(E + m)}\\
0
\end{array}} \right), \nonumber \\[2pt]
{u_ - }({k_z}) & = & \sqrt {\frac{{E + m}}{2}} \left( {\begin{array}{*{20}{c}}
0\\
1\\
0\\
{ - {k_z}/(E + m)}
\end{array}} \right),
\label{eqspinors}
\end{eqnarray}
in the original Dirac basis.  Here $ E \equiv ( \boldsymbol k^2 + {m^2})^{1/2} $  and the spinors have been normalized as in 
ref.\ \cite{nagashima2013}. Fermions with spinors  ${u_ + }({k_z})$  or ${u_ - }({k_z})$  are customarily referred to as being right-handed or left-handed in the relativistic limit where they are in an approximate eigenstate of the chirality operator $\gamma^5$.

Figure \ref{figure 3} shows the effects of the projection operators ${P_L}$  and ${P_L}'$   acting on a fermion with negative helicity (red) or positive helicity (blue), where the normalized projections 
${N_L} \equiv |{P_L}{u_ \pm }{|^2}/|{u_ \pm }{|^2}$  and ${N_L}' \equiv |{P_L}'{u_ \pm }{|^2}/|{u_ \pm }{|^2}$  are plotted as a function of ${k_z}/m.$  It can be seen that the  operator $\slashed{k}/m$  from the Feynman propagator converts a left-handed projection operator into a right-handed projection operator, in agreement with eq.\ \eqref{eqconvert}.  Similar results (not shown) apply to the right-handed projection operator.  The results shown in figure \ref{figure 3} were calculated numerically using the spinors of eq.\ \eqref{eqspinors} and the Dirac matrices, and they are not dependent on the use of any identities such as that of eq.\ \eqref{eqgam5}. 

In evaluating scattering amplitudes, it is commonly assumed that 
$(1 - {\gamma ^5}){u_ + }(k) \approx 0$  and can be neglected in the relativistic limit of $v \to c.$   Eq.\ \eqref{eqconvert} shows that this approximation is not valid when used in conjunction with the divergent tensor product in the propagator.  The use of this approximation may explain why the charge-changing weak interactions of right-handed fermions do not appear to have been discussed previously.

\section{Origin of the enhanced interaction}

The physical origin of the enhanced scattering amplitude for right-handed fermions will be discussed in this section.  This raises a number of issues that provide part of the motivation for the alternative quantization approach considered in section 5.

\subsection{Divergent propagator and unitarity}

As mentioned previously, these effects are possible because a massive “right-handed” fermion will always have a vanishingly-small left-handed component even in the relativistic limit, which can be multiplied by the divergent off-diagonal terms in the propagator to give a large but finite contribution to the scattering cross section.   

The Feynman propagator corresponds to the probability amplitude to create a particle at one location and then annihilate it at another location.  As a result, unitarity places a bound on the propagator for a scalar field that is given by
\begin{equation}
\int {{d^4}\boldsymbol{r}} |\Delta (r){|^2} \le 1.
\label{eqbounda}
\end{equation}	  
In momentum space, the corresponding bound is given by
\begin{equation}
\int \frac { {d^4}\boldsymbol{p}} {(2 \pi )^4} |\Delta (p){|^2} \le 1.
\label{eqboundb}
\end{equation}

The divergent nature of the 
${q_\mu }{q_\nu }/{M_W}^2$  term in the propagator for the W boson makes it impossible to satisfy the unitarity bounds of equations \eqref{eqbounda} and \eqref{eqboundb} at sufficiently high energies.  This is consistent with the fact that an infinitesimal probability amplitude can be multiplied by a divergent factor in the propagator to dominate the scattering cross section.

Nevertheless, the Standard Model is constructed in such a way as to be unitary.  In this case, it can be seen that there is a great deal of cancellation in eq.\ \eqref{eqconvert}, since the factor of $\slashed{k}$  on the left-hand side of the equation diverges in the limit of high energies while the right-hand side of the equation does not.  This cancellation plays a role in maintaining unitarity despite the divergent nature of the ${q_\mu }{q_\nu }/{M_W}^2$  term in the propagator.

The fact that an infinitesimal probability amplitude can dominate the scattering cross section while the Feynman propagator does not satisfy the usual bounds from unitarity  provides part of the motivation for considering the alternative approach discussed in section 5.

\subsection{Rotational symmetry and spin}

\begin{figure}[tbp]
\centering
\includegraphics[width=0.60\textwidth]
{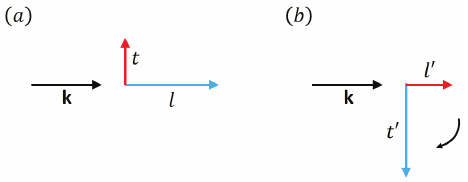}
\qquad
\caption{ Lack of internal rotational symmetry of the field obtained using canonical quantization.  (a)  The field of eq.\ \eqref{eqVmu} has a longitudinal component $l$  that is larger than the transverse component $t$  for large momentum  
$\boldsymbol{k}.$  (b) A  ${90^0}$ rotation of the field components will give a transverse component $t'$   that is larger than the longitudinal component $l'$,  which is inconsistent with the original form of the field.  The field is not symmetric under internal rotations and it can be shown that the action is not invariant as well.  This suggests that the spin may not be conserved.}
\label{figure 4}
\end{figure}

The origin of these effects can be further understood by considering the rotational symmetry of the field of the W  boson after canonical quantization.  In the absence of any interactions, the intrinsic angular momentum (spin) of a particle will be conserved if the system is symmetric under spatial rotations of the internal degrees of freedom of the field \cite{noether1918,schiff1955}.  That is not the case for the field of the W  boson in the Standard Model after canonical quantization, as illustrated in figure \ref{figure 4}.  This suggests that the spin of a  W boson may not be conserved, which provides an intuitive explanation for how the Feynman propagator can convert a left-handed projection operator into a right-handed projection operator.

For an arbitrary field ${\phi _n}(\boldsymbol{k}),$  the irreducible representations of the Lorentz group determine the form of the spin operators 
${\sigma ^i},$ which are the generators of infinitesimal rotations of the components of the field labeled by the index $n$  with the momentum 
$\boldsymbol{k}$  held fixed. (Changes in $\boldsymbol{k}$   can be generated by the boost operator or the orbital angular momentum operator.)  From Noether’s theorem \cite{noether1918}, the spin of a particle will be conserved if the equations of motion are unchanged by the transformation that the ${\sigma ^i}$  generate.

In the canonical quantization of a massive vector field, the longitudinal component of the field for a relativistic particle is larger than the transverse components due to the Lorentz transformation of eq.\ \eqref{eqlambdamup}.  The longitudinal and transverse components are interchanged under a ${90^0}$  rotation of the internal degrees of freedom as shown in figure \ref{figure 4}, which gives a new longitudinal component that is smaller than the transverse component.  Thus the field is not symmetric under a rotation of its internal degrees of freedom and it can be shown that the action is not invariant as well.  This suggests that the spin of a  W  boson may not be conserved. 

The field is symmetric under a complete rotation in which the momentum $\boldsymbol{k}$  and the internal degrees of freedom are all rotated by the same amount, which means that the total angular momentum is conserved.  Thus the change in the spin is compensated by a change in the orbital angular momentum.   

Since the Feynman propagator corresponds to the probability amplitude to create a particle at one location and then annihilate it at another location, the off-diagonal terms in the propagator can be interpreted as a change in the polarization state of a virtual W  boson and thus its spin as it propagates  from one vertex to another.  This is consistent with the argument above based on the lack of rotational invariance.  

\begin{figure}[tbp]
\centering
\includegraphics[width=0.47\textwidth]{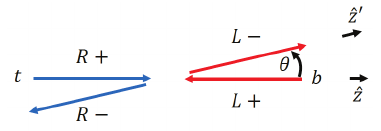}
\qquad
\caption{Lowest-order scattering of a right-handed top quark  t (blue arrows) and a left-handed bottom quark b  (red arrows) as allowed by the tensor product in the Feynman propagator for the W  boson (not shown).  The corresponding spins along the 
$\hat z$  or $\hat z'$   axes are indicated by the 
$ \pm $  signs.  Both of the incoming quarks have spin up along the  $\hat z$ axis while both of the outgoing quarks have spin down along the 
$\hat z'$  axis in the relativistic limit.  This can be interpreted as being due to a change in the spin of the virtual W  boson due to the lack of rotational symmetry of the field. }
\label{figure 5}
\end{figure}

As an example, consider the scattering of a top quark and a bottom quark in the relativistic limit as illustrated in figure \ref{figure 5}, where there is a small angle $\theta $  between the momentum of the incoming top quark and that of the outgoing bottom quark in the center of mass frame.   The contribution from 
${\mathcal{M}_2}$ in eq.\ \eqref{eqM2a} allows the top quark to be right-handed while the bottom quark is left-handed, as illustrated by the notation L and R in the figure.

The $\hat z$  axis is chosen to be towards the right in  figure \ref{figure 5}  while the $\hat z'$  axis is rotated through an angle  $\theta  <  < 1.$ A right-handed fermion has a spin of $ + \hbar /2$  along its direction of travel and a spin of $ - \hbar /2$  in the opposite direction, while the situation is reversed for a left-handed fermion.  In the initial state, this means that the right-handed top quark has a spin of $ + \hbar /2$  along the $\hat z$  axis while the left-handed bottom quark has a spin of $ + \hbar /2$    along the $\hat z$  axis as well.  In the final state, the right-handed top quark has a spin of $ - \hbar /2$  along the $\hat z'$  axis while the left-handed bottom quark also has a spin of  $ - \hbar /2$ along the $\hat z'$    axis.  Thus both of the incoming quarks have spin up while both of the outgoing quarks have spin down along the $\hat z$  or  $\hat z'$   axes.  

This process would correspond to a change in the total spin of the two fermions by $ - 2\hbar $  if it could occur for $\theta  = 0,$  where $\hat z$  and  $\hat z'$ coincide.  As shown in appendix A.2, the scattering amplitude for an event of this kind is proportional to ${sin ^2}(\theta /2),$ which avoids the case where $\hat z' = \hat z.$  But this process can occur for $0 < \theta  <  < 1,$  and when it does, the spins of both outgoing quarks could be measured (in principle) along the original  $\hat z$ axis.  For $\theta  <  < 1,$  both outgoing particles would be found with high probability to have a spin of $ - \hbar /2$   along the $\hat z$  axis, despite the fact that both of the incoming particles had a spin of $ + \hbar /2$  in that direction.

The change in spin of the two fermions in this example can be interpreted as being due to a change in the spin of the virtual W  boson as it propagates between the two vertices, as suggested by the lack of rotational symmetry discussed above.

\subsection{Non-conservation of the weak current}

Although the Standard Model is gauge invariant, the Lagrangian for the Proca equation \cite{greiner1996,griffiths2008,ruegg2004,proca1936} by itself is not gauge invariant due to the mass term in  eq.\ \eqref{eqL}.  As a result, the weak current is not conserved and  ${\partial _\mu }{j^\mu } \ne 0,$  as shown in ref.\ \cite{paschos2007} and appendix A.1.  All of the effects of interest in this paper are dependent on the fact that the weak current is not conserved.

In quantum electrodynamics, the Lagrangian is gauge invariant and  ${\partial _\mu }{j^\mu } = 0.$  The Feynman propagator contains a term proportional to ${q_\mu }{q_\nu }/{q^2}$  in the Landau gauge, which has off-diagonal terms similar to those of the W  boson.  But it follows from gauge invariance that the  ${q_\mu }{q_\nu }/{q^2}$    term has no observable effects, since such a term does not appear in the equivalent Lorentz gauge.  It can be explicitly shown that the effects of the  ${q_\mu }{q_\nu }/{q^2}$    term  must cancel out in quantum electrodynamics by using the Ward identity, which is also based on gauge invariance \cite{peskin1995}.  

In contrast, the effects of the ${q_\mu }{q_\nu }/{M_W}^2$  term for the W  boson propagator need not cancel out because the Lagrangian of eq.\ \eqref{eqL} by itself is not gauge invariant  and the Ward identity does not apply.  It can be shown that the effects of those terms would cancel out if it were the case that  ${\partial _\mu }{j^\mu } = 0.$

As discussed in section 2, the covariant quantization of the field of the W boson is based on the assumption that  that  ${\partial _\mu }{j^\mu } = 0$    \cite{weinberg1995,greiner1996}. The Lorentz condition ${\partial _\mu }{V^\mu } = 0$  can then be derived from the Euler-Lagrange equations.  Thus the canonical quantization of the field is logically inconsistent given that
${\partial _\mu }{j^\mu } \ne 0$  for the weak current.

 It can be shown \cite{greiner1996} from the Proca equation that
\begin{equation}
    \partial _ \mu V ^ \mu =\frac{1}{M ^ 2} \partial _ \mu j ^ \mu, 
    \label{eqdivv}
\end{equation}
while the covariant quantization assumes that 
${\partial _\mu }{V^\mu } = 0.$ This shows that the classical theory has 4 independent polarizations for ${\partial _\mu }{j^\mu } \neq 0$ while the quantized theory only has 3.  The field after canonical quantization is not equivalent to the classical field.

The plane-wave states of equations \eqref{eqVmu} and \eqref{eqlambdamu} do not form a complete set for the expansion of an arbitrary state of the field, only for those with ${\partial _\mu }{V^\mu } = 0.$  A similar situation occurs when using the Coulomb gauge in quantum electrodynamics, where the Coulomb field is not quantized and its effects must be included in some other way.  In contrast, any effects of the time-like component of the field, such as virtual particles generated by a current with ${\partial _\mu }{j^\mu } \ne 0$, are not included in the  canonical quantization of the field of the W  boson in the unitary gauge.  This difficulty can be avoided by quantizing all four components of the field, as described in the next section.

\section{Alternative quantization approach}

An alternative approach that quantizes all four components of the field of a massive vector boson in the unitary gauge will now be considered.  This approach maintains the rotational symmetry of the system and eliminates the ${q_\mu }{q_\nu }/{M_W}^2$  term in the propagator, which gives charge-changing weak interactions only for left-handed fermions.

\subsection{Use of the indefinite metric}

The alternative approach is based on an analogy with the covariant quantization of the electromagnetic field ${A^\mu }$  using the indefinite metric.  All four polarization components are included: two transverse, one longitudinal, and one time-like, as first suggested by Gupta \cite{gupta1950,gupta1957,bleuler1950,gupta1977,cohen-tannoudji1989,itzykson1980}.  The indefinite metric ensures that the longitudinal and time-like photons are not observable in a freely-propagating (radiative) field.  Nevertheless, the time-like and longitudinal photons are an essential part of the theory.  For example, virtual time-like photons associated with the scalar potential ${A^0}$  are necessary to describe the static Coulomb potential and they are responsible for low-energy scattering. 

This suggests an alternative quantization approach in which all four components of the field  
${V^\mu }$ of a massive vector boson are quantized in the unitary gauge using the indefinite metric.  In the rest frame of the particle, an additional polarization mode given by 
\begin{equation}
{\lambda ^\mu }(0,0) = \left( {\begin{array}{*{20}{c}}
1\\
0\\
0\\
0
\end{array}} \right)
\label{eqlambda00}
\end{equation}
is included in addition to the three spatial modes of eq.\ \eqref{eqlambdamu}.  Here  $s = 0$ denotes the time-like component.  

The new time-like polarization mode is nonphysical. As in quantum optics \cite{cohen-tannoudji1989,franson2011} and quantum electrodynamics \cite{gupta1950,gupta1957,bleuler1950,gupta1977}, this requires that the commutation relations for the creation and annihilation operators 
${b^\dag }(0,s)$  and ${b }(0,s)$ in the rest frame have the form
\begin{eqnarray}
[b(0,s),{b^\dag }(0,s')] & = & {\delta _{ss'}}\qquad(s,s' = 1,3) \nonumber\\[2pt]
[b(0,0),{b^\dag }(0,0)] & = &  - 1.\ 
\label{eqcommutatorind}
\end{eqnarray}
Periodic boundary conditions have been used here for simplicity. 

The field operator can then be taken to have the form
\begin{equation}
{V^\mu }(x) = \frac{1}{{{{\left( {2\pi } \right)}^3}}}\sum\limits_{s = 0}^3 {\int {\left( {\frac{{{d^3}\boldsymbol{p}}}{{\sqrt {2{p^0}} }}} \right)} } {e^\mu }(s)b(\boldsymbol{p},s){e^{ - ip \cdot x}} + h.c. \label{eqfield}
\end{equation}
Here ${e^\mu }(s)$ is a fixed set of unit vectors given by ${e^\mu }(s) = {\lambda ^\mu }(0,s)$  as in equations \eqref{eqlambdamu} and \eqref{eqlambda00}.  It will be assumed that the energy of a particle is given by 
$E = {({\boldsymbol{p}^2} + {M^2})^{1/2}},$  which is required for the factor of $\exp (- ip \cdot x)$  in the field operator to be covariant. 

For a classical field, $b(\boldsymbol{p},s)$  would correspond to the Fourier component of the field in the direction specified by  ${e^\mu }(s).$  Those components would transform as a 4-vector under a Lorentz transformation.  For the quantized field, a boost from the rest frame to a coordinate frame with momentum $\boldsymbol{p}$  requires that the annihilation operators also transform as a 4-vector:
\begin{eqnarray}
b(\boldsymbol{p},3) & = & \gamma b(0,3) - \beta \gamma b(0,0), \nonumber \\[2pt]
b(\boldsymbol{p},0) & = & \gamma b(0,0) - \beta \gamma b(0,3).
\label{eqcommutatorsindb}
\end{eqnarray}
Here the z-axis has been chosen in the direction of 
$\boldsymbol{p}$  for simplicity, $\beta $  is the relative velocity between the two coordinate frames, and 
$\gamma  = {(1 - {\beta ^2})^{ - 1/2}}.$   

By combining equations \eqref{eqcommutatorind} and \eqref{eqcommutatorsindb}, it can be shown that
\begin{eqnarray}
[b(\boldsymbol{p},s),{b^\dag }(\boldsymbol{p},s')] & = & {\delta _{ss'}}\qquad(s,s' = 1,3) \nonumber\\[2pt]
[b(\boldsymbol{p},0),{b^\dag }(\boldsymbol{p},0)] 
& =  &  - 1.\ 
\label{eqcommutatorindp}
\end{eqnarray}
Eq.\eqref{eqcommutatorindp} shows that the commutation relations are invariant under a Lorentz transformation, even though the annihilation operators themselves transform as a 4-vector.  This is due to cancellation between the longitudinal and time-like components. It can be shown in the same way that the commutation relations are invariant under spatial rotations as well.  The observable properties of the system are determined by the field equation and the commutation relations, which are symmetric under spatial rotations  due to the unit vectors in eq\ \eqref{eqfield}.

If the field ${\pi ^\mu }$  is defined by 
${\pi ^\mu } \equiv {\dot V^\mu }$, then for the free field
\begin{equation}    
[{V^\mu }(\boldsymbol{x},t),{\pi ^\nu }(\boldsymbol{y},t)] =  - i{\eta ^{\mu \nu }}{\delta ^{(3)}}(\boldsymbol{x} - \boldsymbol{y}).
\label{eqcommutatorV}
\end{equation}
The Feynman propagator $\Delta _{\mu \nu }^A $ can be calculated from equations \eqref{eqfield} and \eqref{eqcommutatorindp} using standard techniques \cite{peskin1995}, with the result that

\begin{equation}
\Delta _{\mu \nu }^A =  - i\frac{{{\eta _{\mu \nu }}}}{{{q^2} - {M_W}^2 + i\varepsilon }}.
\label{eqpropA}
\end{equation}
The sign change in the ${\eta _{00}}$  term comes from $[b(\boldsymbol{p},0),{b^\dag }(\boldsymbol{p},0)] =  - 1.$   Here the superscript 
$A$  refers to the fact that this is the Feynman propagator in the unitary gauge derived using the alternative approach.  

It can be seen that the quantization of all four components of the field eliminates the tensor product ${q_\mu }{q_\nu }/{M_W}^2$  that appears in the conventional propagator of eq.\ \eqref{eqpropu}.  The scattering amplitude ${\mathcal{M}_2}$  with the right-handed projection operator is eliminated from the alternative theory, leaving only the left-handed projection operator in ${\mathcal{M}_1},$  and the alternative quantization approach gives charge-changing weak interactions only for left-handed fermions.

The assumptions inherent in this approach are discussed in more detail in appendix A.3, including the effects of  ${\partial _\mu }{j^\mu } \ne 0.$   For simplicity, the unitary gauge with $\xi  \to \infty $  is chosen in order to eliminate the Goldstone boson and other nonphysical particles.  The field is not intended to satisfy the Euler-Lagrange equations.  Instead, the form of the field is determined by the requirement that the system be symmetric under spatial rotations of the components of the field. 

The same propagator can be obtained in the ‘t Hooft-Feynman gauge using that Lagrangian and more conventional methods.  But in that case the nonphysical Goldstone boson will give an interaction with right-handed fermions, as can be seen from eq.\ \eqref{eqVg}.  A more complicated field for the  W boson would be required in order to eliminate those effects and maintain consistency with the alternative approach in the unitary gauge. 

One might ask whether or not the negative norms inherent in the indefinite metric are reasonable.  Feynman ridiculed the use of the indefinite metric on at least one occasion \cite{feynman1973}, although he eventually changed his mind regarding the role of negative quasiprobability distributions \cite{feynman1987}.  The use of the indefinite metric is logically consistent as long as the physical (observable) states of the system have a positive norm \cite{gupta1950,gupta1957,bleuler1950,gupta1977,cohen-tannoudji1989}.  

Aside from eliminating the interaction with right-handed fermions, the alternative quantization approach also ensures that the properties of the W  and ${Z^0}$  bosons become the same as those of a photon (aside from their interactions) in the relativistic limit where the mass of the particle is negligible compared to its total energy, as one might expect.  That is not the case in the Standard Model \cite{weinberg1995}.

\subsection{Feasibility of experimental tests}

The feasibility of experimentally distinguishing between the predictions of the alternative quantization approach and those of the Standard Model will be considered in this section.  The alternative approach appears to be consistent with existing experiments, but additional high-energy experiments may be required in order to distinguish between the two theories.

The largest fermion mass ${m_f}$  in most scattering experiments is much smaller than 
${M_W},$  as is the case for the scattering of an electron and a neutrino as shown in figure  2(a).   The amplitude ${\mathcal{M}_2}$  in the Standard Model allows the scattering of a right-handed electron, but  ${\mathcal{M}_2}$ is proportional to 
${({m_e}/{M_W})^2}$  and the corresponding cross section is proportional to  
${({m_e}/{M_W})^4} \approx 1.63 \times {10^{ - 21}}.$   This is clearly too small to be experimentally observable. The situation is not much better for the scattering of an up quark and a down quark, where ${({m_d}/{M_W})^4} \approx 1.27 \times {10^{ - 17}}.$  Experimental searches \cite{wu1957,quin1989,ashery2017} for right-handed interactions (beyond the Standard Model) have given negative results, but the effects of interest here are too small to have been observed in those experiments. 

\begin{figure}[tbp]
\centering
\includegraphics[width=0.27\textwidth]
{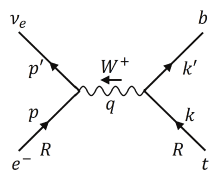}
\qquad
\caption{ Lowest-order Feynman diagram for the scattering of a top quark and an electron with the exchange of a ${W^ + }$  boson.  The tensor product in the propagator allows this process to occur for a right-handed electron and a right-handed top quark, as indicated by the letter R.  The Standard Model predicts a scattering cross section that is proportional to  ${({m_e}/{M_W})^2},$    which is too small to distinguish between the two theories. }
\label{figure 6}
\end{figure}

These examples suggest that any observable difference between the two theories would require experiments involving the top quark.  For example, figure \ref{figure 6} shows the Feynman diagram for the scattering of a top quark and an electron.  In this case, the right-handed cross section from 
$|{\mathcal{M}_2}{|^2}$ is proportional to ${m_e}^2{m_t}^2/{M_W}^4 \approx 1.87 \times {10^{ - 10}},$  which is still too small to be experimentally observable.  The analogous scattering of a top quark and a down quark would be proportional to  
${m_d}^2{m_t}^2/{M_W}^4 \approx 1.65 \times {10^{ - 8}}.$

\begin{figure}[tbp]
\centering
\includegraphics[width=0.7\textwidth]{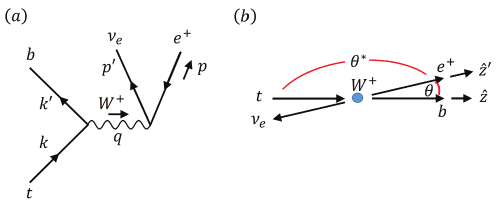}
\qquad
\caption{ (a) Feynman diagram for the decay of a top quark t into a bottom quark b and a ${W^ + }$   boson, which can then decay into a positron 
${e^ + }$  (or other lepton) and a neutrino  
${\nu _e}.$  (b)  Definition of the angles $\theta $  and $\theta ^*$   in the rest frame of the  ${W^ + }$   boson.  The differences between the predictions of the alternative quantization approach and the Standard Model are negligible to lowest order. }
\label{figure 7}
\end{figure}

It has been suggested \cite{peskin2017} that  experiments involving the decay of a top quark can provide evidence for the enhancement in the longitudinal component of the W boson  due to the Lorentz transformation of eq.\ \eqref{eqlambdamup}.  This enhancement is closely related to the presence of the  
${q_\mu }{q_\nu }/{M_W}^2$ term in the Feynman propagator and it does not occur in the alternative quantization approach.  As illustrated in figure \ref{figure 7}, a top quark can decay into a bottom quark and a
${W^ + }$ boson, which can subsequently decay into a lepton and a neutrino \cite{aaltonen2010,aaltonen2013,khachatryan2016,kane1992,aguilar2007,czarnecki2010}.  The angles 
$\theta $  and $\theta ^*$  are defined in the rest frame of the ${W^ + }$  boson as illustrated in figure 7(b). The measured angular distribution of 
$\theta ^*$ is related to the polarization state (helicity) of the  ${W^ +.}$

\begin{figure}[tbp]
\centering
\includegraphics[width=0.50\textwidth]
{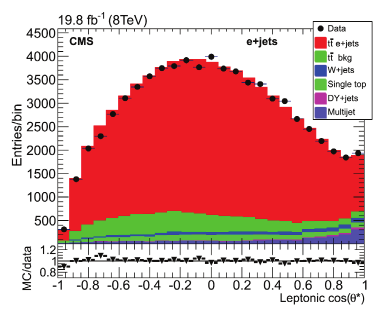}
\qquad
\caption{ Experimental measurement of the angular distribution of the secondary positrons from the decay of a top quark as measured at the CMS detector at the LHC \cite{khachatryan2016}.  The theoretical predictions are shown in red.  Figure reproduced from ref.\ \cite{khachatryan2016}. }
\label{figure 8}
\end{figure}

The lowest-order amplitude $ \mathcal{M}_1'$  for this process due to the ${\eta _{\mu \nu }}$  term in the propagator is the same in the alternative approach as it is in the Standard Model. The amplitude $\mathcal{M}_2'$  for the Feynman diagram of figure 7(a) in the Standard Model is given by
\begin{eqnarray}
{M_2}' & \approx &  - ig{'^2}\left( {\frac{{{m_t}{m_e}}}{{{M_W}^2}}} \right)\left[ {{{\bar u}_\nu }(p')(1 + {\gamma ^5}){v_{e + }}(p)} \right] \nonumber\\[2pt]
& \times &  \left( {\frac{1}{{{q^2} - {M_W}^2 + i{M_W}\Gamma }}} \right)\left[ {{{\bar u}_b}(k')(1 + {\gamma ^5}){u_t}(k)} \right],
\label{eqM2p}
\end{eqnarray} 
while $ \mathcal{M}_2' = 0 $  in the alternative approach.  Here $ \Gamma $  is the decay width of the $ { \text{W}^ + } $  resonance.  Once again, the probability of an event from $ \mathcal{M}_2' $  is proportional to $ {({m_e}/{M_W})^2} <  < 1, $   and the differences between the predictions of the Standard Model \cite{kane1992,aguilar2007,czarnecki2010} and the alternative approach are negligible to lowest order for experiments of this kind, as shown in appendix A.4.  The experimentally-observed angular distribution is shown in figure \ref{figure 8} along with the theoretical prediction \cite{khachatryan2016}.  

This suggests that the top quark must appear in both vertices in order to experimentally distinguish between the two theories.  For example, the predicted cross section for the scattering of a right-handed top quark and a left-handed bottom quark, as illustrated in figures 1 and 2(b), is proportional to ${({m_t}/{m_W})^4} = 21.4$  in the Standard Model, which is an order of magnitude larger than the cross section for a left-handed top quark.  Unfortunately, the direct observation of the scattering of top and bottom quarks is complicated by a number of factors, including the confinement of the quarks and the contribution from the strong force, and there does not appear to be any experimental data of that kind to date.

\begin{figure}[tbp]
\centering
\includegraphics[width=0.40\textwidth]
{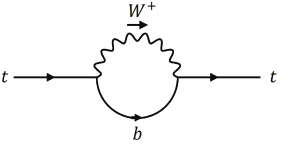}
\qquad
\caption{ Higher-order correction to the self-energy of the top quark.  Here a top quark t spontaneously decays into a virtual state containing a bottom quark b and a ${W^ + }$  boson,  which subsequently annihilate to restore the top quark.  Higher-order corrections of this kind may be able to distinguish between the predictions of the alternative quantization approach and those of the Standard Model in high-precision experiments.    }
\label{figure 9}
\end{figure}

Feynman diagrams with a top quark at both vertices can also occur in higher-order corrections, as is illustrated in figure \ref{figure 9}.  Here a top quark is annihilated to produce a virtual bottom quark along with a ${W^ + }$ boson, both of which are subsequently annihilated to restore the original top quark.  This process is analogous to the self-energy of an electron in quantum electrodynamics. Higher-order effects of various kinds can produce small changes in the predicted scattering cross sections, production rates, and decay rates.  The analysis of higher-order effects at high energies is difficult due to the large number of particles in the Standard Model and the correspondingly large number of Feynman diagrams.

A comparison of the higher-order corrections from the alternative quantization approach with those from the Standard Model is beyond the intended scope of this paper.  It may be worth noting, however, that the observed production rate of 
$t\bar tW$ events at the LHC is two standard deviations larger than would be expected from the Standard Model \cite{cms2022,atlas2023}.

\section{Summary and conclusions}

The Lagrangian for the charge-changing weak interaction is proportional to the left-handed projection operator 
$(1 - {\gamma ^5})/2.$   Nevertheless, it was shown here that the Standard Model predicts charge-changing cross sections  for right-handed fermions that can be larger than those for left-handed fermions if the mass is sufficiently large, as is the case for the top quark.  Here we are using the conventional terminology in which a massive fermion with its spin parallel to its momentum is referred to as being right-handed in the relativistic limit, where it is in an approximate eigenstate of the chirality operator.

These effects are due to the way in which the field of the W  boson  is quantized, which gives a divergent tensor product ${q_\mu }{q_\nu }/{M_W}^2$  in the Feynman propagator in the unitary gauge.  The tensor product in the propagator was used to rewrite the conventional expression for the scattering amplitude in an equivalent form that includes the right-handed projection operator 
$(1 + {\gamma ^5})/2,$  which allows interactions with right-handed fermions to occur.  Mathematically, this can be understood from the fact that the $\slashed{k}/m$  operator from the Feynman propagator can convert a left-handed projection operator into a right-handed projection operator (and vice versa), as shown in eq.\ \eqref{eqconvert} and illustrated in figure \ref{figure 3}.

These effects are possible because a massive “right-handed” fermion will always contain a vanishingly-small left-handed component even in the relativistic limit, which is multiplied by the divergent cross-terms in the propagator to give a large but finite contribution to the cross section.  This is not merely a semantic issue, however, since experiments to date have shown charge-changing weak interactions only for left-handed fermions, and it is an open experimental question as to whether or not right-handed fermions can have larger scattering cross sections than left-handed fermions at sufficiently high energies.  

One might suspect that Feynman diagrams involving other particles in the Standard Model might cancel the effects obtained here.  But to lowest order (tree level), there is no contribution from other particles such as the Higgs boson in the unitary gauge.  The same results can be obtained using the ‘t Hooft-Feynman gauge, where the vertex factor for the nonphysical Goldstone boson explicitly includes a term that is proportional to the right-handed projection operator  $(1 + {\gamma ^5})/2.$

Although the Standard Model is unitary and its predictions are well-defined, there are a number of issues that suggest that it may be worthwhile to consider an alternative quantization approach for the charge-changing weak interaction:
\begin{itemize}
    \item
    The canonical quantization of the field of the W boson in the unitary gauge is based on the assumption that ${\partial _\mu }{j^\mu } = 0$  \cite{weinberg1995,greiner1996} whereas the weak current is not conserved \cite{paschos2007}, which is logically inconsistent.
    \item
    The quantized theory is inconsistent with the classical theory (Proca equation), which does not assume that ${\partial _\mu }{j^\mu } = 0$  or ${\partial _\mu }{A^\mu } = 0.$
    \item 
    The lack of rotational symmetry of the internal degrees of freedom of the field after canonical quantization suggests that the spin of a  W boson may not be conserved.  This provides an intuitive explanation for how its propagator can convert a left-handed projection operator into a right-handed projection operator.
    \item 
    The divergent nature of the Feynman propagator allows an infinitesimal probability amplitude to dominate the scattering cross section at high energies, and the propagator does not satisfy the usual bounds from unitarity.
    \item 
    Existing experiments have not demonstrated charge-changing weak interactions for right-handed fermions, although additional high-energy experiments may be required in order to observe these effects.    
\end{itemize}

The presence of the tensor product in the conventional propagator is due to the fact that only 3 of the 4 components of the field of a massive vector boson are quantized in the unitary gauge of the Standard Model \cite{weinberg1995}.  An alternative approach was considered here in which all 4 components of the field are quantized in a covariant way using the indefinite metric, which maintains the rotational symmetry of the system and ensures that the interaction with right-handed fermions is negligible.   The properties of the  W and ${Z^0}$  vector bosons become the same as those of a photon (aside from their interactions) in the relativistic limit where the mass of the particle is negligible compared to its total energy, as one might expect.

These effects appear to be experimentally observable only for interactions involving the top quark due to its large mass.  Higher-order corrections, such as the one shown in figure \ref{figure 9}, appear to be the most likely source of experimentally-observable differences between the alternative quantization approach and the Standard Model. These results may have an impact on the interpretation of high-energy experiments that are intended to search for new physics beyond the Standard Model.

In summary, the Standard Model predicts that right-handed fermions can have larger charge-changing weak interactions than left-handed fermions for particles with a sufficiently large mass, such as the top quark.  Additional high-energy experiments may be required in order to distinguish between the predictions of the Standard Model and those of the alternative quantization approach considered here.

\appendix

\section{Appendices}

\subsection{Properties of the weak current}

The Lagrangian of eq.\ \eqref{eqL} is not gauge invariant by itself due to the presence of the mass term and the weak current is not conserved as a result \cite{paschos2007}, which plays an essential role in the origin of the effects of interest here.   A simplified proof that ${\partial _\mu }{j^\mu } \ne 0$  is given in this appendix. 

Aside from a possible constant, the charged weak current has the form
\begin{equation}
{j^\mu }(x) = {\bar \psi _u}(x){\gamma ^\mu }(1 - {\gamma ^5}){\psi _d}.
\label{eqjmu}
\end{equation}
Here the subscripts $u$  and $d$  refer to a pair of up-type and down-type fermions, such as an electron and a neutrino, or a top quark and a bottom quark.  In k-space, this becomes
\begin{equation}
j ^\mu(k) = {\bar u_u}({k_u}){\gamma ^\mu }(1 - {\gamma ^5}){u_d}({k_d}),
\label{eqjk}
\end{equation}
with  $k = {k_u} - {k_d}.$ 

For simplicity, consider a spinor ${u_d}$  that is an approximate eigenstate of $(1 - {\gamma ^5})/2.$  The divergence of eq.\ \eqref{eqjk} is then given by
\begin{equation}
{\partial _\mu }{j^\mu }(k) = 2i{\bar u_u}({k_u})({k_{u\mu }} - {k_{d\mu }}){\gamma ^\mu }{u_d}({k_d}),
\label{eqdiva}
\end{equation}
since $(1 - {\gamma ^5}){u_d} \approx 2{u_d}.$  Using the fact that 
$\slashed{k}{_d}{u_d}({k_d}) = {m_d}{u_d}({k_d})$  as in the text, along with a similar expression for 
${\bar u_u}({k_u}),$  eq.\ \eqref{eqdiva} reduces to 
\begin{equation}
{\partial _\mu }{j^\mu }(k) = 2i({m_u} - {m_d}){\bar u_u}({k_u}){u_d}({k_d}).
\label{eqdivb}
\end{equation}

The two masses are not equal and the inner product of the two spinors is not zero in general.  Thus eq.\ \eqref{eqdivb} shows that
${\partial _\mu }{j^\mu } \ne 0$   and the charged weak current is not conserved.  The same is true for more general spinors ${u_d}$ that are not eigenstates of $(1 - {\gamma ^5})/2$  \cite{paschos2007}.  

It can be shown that the effects of the off-diagonal terms in the conventional Feynman propagator of eq.\ \eqref{eqpropu} would cancel out unless  ${\partial _\mu }{j^\mu } \ne 0.$  This is equivalent to the fact that the Ward identity does not apply due to the lack of gauge invariance of eq.\ \eqref{eqL}.

\subsection{Angular distribution of the scattering of a top and bottom quark}

This appendix calculates the angular dependence of the scattering amplitude for the process shown in figure \ref{figure 5}, where the total spin of a top quark and a bottom quark changes by $ - 2\hbar $  in the final state for small values of the angle $\theta .$ 

Eq.\ \eqref{eqM2a} in the text gives the lowest-order scattering amplitude $\mathcal{M} _2 $ due to the tensor product in the Feynman propagator in the unitary gauge of the Standard Model.  The right-hand side of the equation contains a product of factors $ {F_R '} $  given by
\begin{equation}
{F_R '} \equiv {\bar u_b}(k')(1 + {\gamma ^5}){u_t}(k),
\label{eqFRB}
\end{equation}
where the momenta are defined in figure 2(b).  The incoming top quark is right-handed and has spin up along the $\hat z$  axis so that 
\begin{equation}
{u_t}(k) = {c_t}\left( {\begin{array}{c}
1\\
0\\
1\\
0
\end{array}  }
\right).
\label{equt2}
\end{equation}
Here the relativistic limit has been assumed for both particles.  

The outgoing bottom quark is left-handed and has spin down along the  $\hat z'$ axis.  If the angle 
$\theta $  were zero, the bottom quark would have a spinor ${u_{b0}}(k')$ given by
\begin{equation}
{u_{b0}}(k') = {c_b}\left( {\begin{array}{*{20}{c}}
0\\
1\\
0\\
{ - 1}
\end{array}} \right).
\label{eqb0}
\end{equation}
For $\theta  \ne 0,$  a rotation about the
$\hat x$ axis gives \cite{baym1973}
\begin{eqnarray}
{u_b}(k') & = & {e^{-i \boldsymbol{\sigma } \cdot \,\boldsymbol{\theta }/2}}{u_{b0}}(k') \nonumber \\[2pt]
 & = & \left[ {\cos \left( {\frac{\theta }{2}} \right) - i{\sigma _x}\sin \left( {\frac{\theta }{2}} \right)} \right]{u_{b0}}(k').
\label{equbsin}
\end{eqnarray}
Here
\begin{equation}
\boldsymbol{\sigma } = \left( {\begin{array}{*{20}{c}}
\boldsymbol{\tau }&0\\
0&\boldsymbol{\tau }
\end{array}} \right),
\end{equation}
where $\boldsymbol{\tau }$  represents the Pauli spin matrices \cite{baym1973,bjorken1964}. 

Inserting these expressions into eq.\ \eqref{eqFRB} gives
\begin{eqnarray}
{F_R '} & = & {c_t}{c_b}{\left( {\begin{array}{*{20}{c}}
0\\
1\\
0\\
{ - 1}
\end{array}} \right)^\dag }{\left[ {\cos \left( {\frac{\theta }{2}} \right) - i{\sigma _x}\sin \left( {\frac{\theta }{2}} \right)} \right]^\dag }{\gamma ^0}\left( {1 + {\gamma ^5}} \right)\left( {\begin{array}{*{20}{c}}
1\\
0\\
1\\
0
\end{array}} \right) \nonumber \\[2pt]
& = & 4i{c_t}{c_b}\sin \left( {\frac{\theta }{2}} \right).
\label{eqFRd}
\end{eqnarray}
Similar results can be obtained for an analogous factor of
${F_L '}$  on the left-hand side of eq.\ \eqref{eqM2a}, with the result that
\begin{equation}
{\boldsymbol{M}_2} =- 16i{c_t}^2{c_b}^2{\sin ^2}\left( {\frac{\theta }{2}} \right)\left( {\frac{{{m_t}^2}}{{{M_W}^2}}} \right)\frac{{g{'^2}}}{{{q^2} - {M_W}^2}}.\label{eqM2c}
\end{equation}

It can be seen that the Standard Model predicts a probability amplitude for this process that is proportional to ${\sin ^2}(\theta /2)$  and a probability that is proportional to 
${\sin ^4}(\theta /2)$.  Both of the incoming particles have spin up in the relativistic limit while both of the outgoing particles have spin down with high probability along the z axis for
$\theta  <  < 1,$   giving a  change in the total spin of $ - 2\hbar .$    A scattering process of this kind does not occur in the alternative quantization approach, where the amplitude
${\mathcal{M}_2}$  is zero.

As discussed in the text, the change in spin of
$ - 2\hbar $  in the final state can be interpreted as being due to a change in spin of the virtual W  boson as it propagates from one vertex to another, as suggested by the lack of rotational invariance of the internal degrees of freedom of the field.

\subsection{Fundamental assumptions in the alternative quantization approach}

This appendix discusses the fundamental assumptions inherent in the alternative quantization approach based on the use of the indefinite metric \cite{gupta1950,gupta1957,bleuler1950,gupta1977,cohen-tannoudji1989,itzykson1980}.  These assumptions are required in order to maintain the rotational symmetry of the internal degrees of freedom of the field, which ensures that the spin of a  W  boson will be conserved.  The differences between this approach and earlier approaches \cite{itzykson1980} that also quantized all four components of the field will be discussed as well.

The alternative quantization approach is similar in spirit to Chapter 5 of Weinberg’s text \cite{weinberg1995}, where the forms of the field operators are determined from a few fundamental requirements, such as causality.  The Lagrangian does not play a role in that approach.

For simplicity, the unitary gauge will be used in order to eliminate any nonphysical particles.  The alternative approach is based on four fundamental assumptions:
\begin{enumerate}
    \item 
    All four components of a massive vector field must be quantized.
    \item 
    The nonphysical nature of the time-like component of the field requires that\\ $[b(0,0),{b^\dag }(0,0)] < 0$  in the rest frame, while the other three commutators are positive.
    \item 
    The field operator must transform as a 4-vector under Lorentz transformations, including spatial rotations.
    \item
    The system must be symmetric under spatial rotations of the components of the field.    
\end{enumerate}

The approach described in the text satisfies this set of requirements.  The same is true for the usual covariant quantization of the electromagnetic field \cite{gupta1950,gupta1957,bleuler1950,gupta1977,cohen-tannoudji1989}, except that there is no rest frame for a photon and an arbitrary reference frame must be chosen instead.  The field in eq.\ \eqref{eqfield} transforms as a 4-vector, given the transformation properties of the annihilation operators 
$ {b}(\boldsymbol{p},s).$

Nevertheless, the commutation relations in eq.\ \eqref{eqcommutatorindp} are invariant under Lorentz transformations and spatial rotations, which is due to cancellations between the longitudinal and time-like components.   It can be seen from the derivation in equations \eqref{eqcommutatorsindb} and \eqref{eqcommutatorindp} that all four commutators must have equal magnitudes in order for this cancellation to occur, as was chosen to be the case in eq.\ \eqref{eqcommutatorind}. 

This approach is equivalent to quantizing a set of independent harmonic oscillators, as is often the case in quantum field theory \cite{peskin1995}. ${V^\mu }$ and  ${\pi ^\mu(k) } \equiv {\dot V^\mu(k) }$ 
correspond to the displacement operator $\hat x$  and the momentum operator $\hat p$  for each oscillator.  The field is not intended to satisfy the Euler-Lagrange equations and ${\pi ^0}$  cannot be calculated in the usual way from the Lagrangian.

The use of the indefinite metric is logically consistent as long as all physical (observable) states of the system have a nonnegative norm.  First consider the situation in quantum electrodynamics, where the Lorentz condition plays an important role in maintaining the consistency of the theory.  It can be shown that the Lorentz condition of eq.\ \eqref{eqlorentz} cannot be satisfied as an operator identity.  This difficulty can be avoided by applying the Lorentz condition to the physical states $\left| \psi  \right\rangle $  of the system rather than the field, which gives \cite{gupta1950,gupta1957,bleuler1950,gupta1977,cohen-tannoudji1989}
\begin{equation}
{\partial _\mu }{A^{( + )\mu }}\left| \psi  \right\rangle  = 0.
\label{eqlpsi}
\end{equation}
Here ${A^{( + )}}$ is the positive-frequency component of the electromagnetic field.  This condition is less restrictive than applying the Lorentz condition to the field itself, since it allows virtual states involving the time-like and longitudinal components of the field.  Eq.\ \eqref{eqlpsi} will be satisfied for all time if it is satisfied initially, provided that
${\partial _\mu }{j^\mu } = 0$  \cite{cohen-tannoudji1989}.

Taking the Fourier transform of eq.\ \eqref{eqlpsi} gives  
$( - i\omega {a_k}^0 + ik{a_k}^l)\left| \psi  \right\rangle  = 0$  in quantum electrodynamics, where $l$  denotes the longitudinal component and 
${a_k}$  is the Fourier transform of the field.  On the mass shell where $\omega  = k,$  this reduces to
\begin{equation}
{a_k}^0\left| \psi  \right\rangle  = {a_k}^l\left| \psi  \right\rangle .\label{eqsuppl}
\end{equation}
This shows that the amplitudes of the longitudinal and time-like components must be equal on the mass shell.  The negative norm associated with the time-like component cancels out the positive norm for the longitudinal component, giving a total norm of zero for the nonphysical states.  This ensures that the longitudinal and time-like photons are unobservable in a freely-propagating field.

For a virtual state,  $\omega $ need not equal
$k$  and eq.\ \eqref{eqsuppl} does not hold in general.  For example, the Coulomb field from a static charge distribution consists entirely of time-like virtual photons and the state has a negative norm.  This is acceptable because virtual states cannot be directly observed.  The expectation value of an observable operator evaluated in a state with a negative norm is discussed in ref.\ \cite{cohen-tannoudji1989}, for example.

Now consider a vector field ${V^\mu }$   with mass M.  If eq.\ \eqref{eqlpsi} still holds, its Fourier transform would give 
$( - i{E_k}{v_k}^0 + ik{v_k}^l)\left| \psi  \right\rangle  = 0,$  where we have used the form of the field in eq.\ \eqref{eqfield} and ${v_k}$  is the Fourier transform of the field.  Thus
\begin{equation}
{v_k}^0\left| \psi  \right\rangle  = \frac{k}{{{E_k}}}{v_k}^l\left| \psi  \right\rangle  \le {v_k}^l\left| \psi  \right\rangle
\label{eqsupplp}
\end{equation}
would be satisfied on the mass shell.  This would give an amplitude for the time-like component that is less than that of the longitudinal component, which would ensure that the norms of the physical states are nonnegative.

But the continuity equation 
${\partial _\mu }{j^\mu } = 0$  is not satisfied for the weak interaction \cite{paschos2007} as shown in appendix A.1, and the Lorentz condition cannot be satisfied in general as a result \cite{jackson1962}.  This is a major problem for the canonical quantization approach, where the form of the field operator in eq.\ \eqref{eqVmu} was determined in part by the assumption that
${\partial _\mu }{j^\mu } = 0.$  Thus the canonical quantization approach is logically inconsistent if ${\partial _\mu }{j^\mu } \ne 0,$  as can be seen from the derivation in ref.\ \cite{weinberg1995}.

This difficulty can be avoided in the alternative approach by making use of the fact that the lifetime of the W  boson is so short that it cannot be directly observed.  As a result, the W  boson can always be considered to be a virtual particle in a larger process, such as the decay of the top quark illustrated in figure \ref{figure 7}.  That avoids any difficulties with the interpretation of negative norms even though 
${\partial _\mu }{j^\mu } \ne 0.$     An example of such a calculation is described in the following appendix.  Treating the W boson as a virtual particle in this way appears to be necessary in order to maintain the rotational symmetry of the field and conserve the spin. 

The Proca equation for a vector particle of mass M  has been quantized previously using all four components of the field and the indefinite metric by adding a gauge-fixing term 
$\mathcal{L}'$  to the Lagrangian \cite{itzykson1980}, where
\begin{equation}
\mathcal{L}'=\frac{{1}}{{2\xi }}{\left( {{\partial _\mu }{V^\mu }} \right)^2}.\label{eqLp}
\end{equation}
In the unitary gauge where $\xi  \to \infty ,$ the commutation relations in that approach are very different from those used here.  In particular, the magnitudes of the commutators are not equal as required by assumption 4 above, and the field is not symmetric under rotations nor is the spin conserved.  As already mentioned, the field is not derived from the Lagrangian in the alternative quantization approach considered here.

The most general form of the Proca equation is \cite{proca1936,nagashima2013,greiner1996,griffiths2008,ruegg2004}
\begin{equation}
{\partial _\mu }\left( {{\partial ^\mu }{V^\nu } - {\partial ^\nu }{V^\mu }} \right) + {M^2}{V^\nu } = {j^\nu }.\label{eqProca1}
\end{equation}
If ${\partial _\mu }{j^\mu } = 0,$  it can be shown that ${\partial _\mu }{V^\mu } = 0$  and eq.\ \eqref{eqProca1} reduces to
\begin{equation}
\left( {{\partial _\mu }{\partial ^\mu } + {M^2}} \right){V^\nu } = {j^\nu }.\label{eqProca2}
\end{equation}
Equations \eqref{eqProca1} and \eqref{eqProca2} are not equivalent if 
${\partial _\mu }{j^\mu } \ne 0,$  as is the case for the weak interaction. Aside from the boundary conditions,  $\Delta _{\mu \nu }^U$ is the Green’s function for eq.\ \eqref{eqProca1} while 
$\Delta _{\mu \nu }^A$ is the Green’s function for eq.\ \eqref{eqProca2}.  In order to conserve the spin, eq.\ \eqref{eqProca2} would have to be the correct field equation for the W boson even when 
${\partial _\mu }{j^\mu } \ne 0.$

\subsection{Angular distribution of the decay of the top quark}

This appendix calculates the angular distribution of the positron created in the decay of a top quark using the alternative quantization approach, as illustrated in figure \ref{figure 7}.  It has been suggested \cite{peskin2017} that the angular distribution is sensitive to the enhancement in the longitudinal component of a relativistic  
 W boson, which does not exist in the alternative quantization approach.  It will be found that the difference between the predictions of the alternative quantization approach and the Standard Model are negligible to lowest order.

Using the propagator $\Delta _{\mu \nu }^A$  from the alternative quantization approach, the lowest-order amplitude $\mathcal{M} _A $  from the Feynman diagram of figure \ref{figure 7} is given by
\begin{equation}
\mathcal{M} _ A = \varepsilon {j_2}^\mu {\eta _{\mu \nu }}{j_1}^\nu ,
\label{eqMd}
\end{equation}
where
\begin{eqnarray}
\varepsilon & = &  - i\frac{{g{'^2}}}{{\left( {{q^2} - {M_W}^2 + i{M_W}\Gamma } \right)}}, \nonumber \\[2pt]
{j_1}^\nu & = & {{\bar u}_b}(k'){\gamma ^\nu }(1 - {\gamma ^5}){u_t}(k), \nonumber \\[2pt]
{j_2}^\mu & = & {{\bar u}_\nu }(p'){\gamma ^\mu }(1 - {\gamma ^5}){v_{{e^ + }}}(p).
\label{eqeps}
\end{eqnarray}

Consider the rest frame of the ${W^ + }$   boson as illustrated in figure 7(b), where $\theta ^ *$  is conventionally defined as the angle between the charged lepton 3-momentum in the ${W^ + }$  rest frame and the ${W^ + }$  momentum in the rest frame of the top quark \cite{aaltonen2013}, while 
$\theta  = \pi  - \theta ^ *.$   The mass of the bottom quark will be neglected as is often done, which gives results that are accurate to approximately $0.2\% $ .  In that case, the bottom quark as well as the neutrino are highly relativistic and they must be left-handed, since the tensor product term does not appear in 
$\Delta _{\mu \nu }^A.$  The positron must be right-handed for the same reason.  

The spinor ${u_b}$  for the bottom quark corresponds to a negative spin along the 
$\hat z$  axis in coordinate frame $O$  in figure 7(b).  The spinors ${v_{{e^ + }}}$  and 
${u_\nu }$  for the positron and  neutrino correspond to positive spins along the 
$\hat z'$ axis in a coordinate frame $O'$  that is rotated through an angle $\theta $  about the y axis with respect to $O.$   The components of the vector ${j_2}'$  in that coordinate frame will be transformed later into the  $O$ coordinate frame. The corresponding spinors in the original Dirac representation are then given by \cite{baym1973,bjorken1964}
\begin{equation}
{u_b} = {c_b}\left( {\begin{array}{*{20}{c}}
0\\
1\\
0\\
{ - 1}
\end{array}} \right)\quad {v_{{e^ + }}} = {c_{{e^ + }}}\left( {\begin{array}{*{20}{c}}
0\\
{ - 1}\\
0\\
1
\end{array}} \right)\quad {u_{_\nu }} = {c_{_\nu }}\left( {\begin{array}{*{20}{c}}
1\\
0\\
{ - 1}\\
0
\end{array}} \right).\label{eqspinorsd}
\end{equation}
Here ${c_b} \equiv \sqrt {({E_b} + {m_b})/2} $  with similar definitions for ${c_{{e^ + }}}, $  
${c_\nu },$ and  ${c_t}.$  

The top quark is assumed to be unpolarized and it can have positive or negative helicity with equal probability.   Its velocity is significantly less than the speed of light and the corresponding spinors ${u_{t + }}$  and   ${u_{t - }}$ are given by
\begin{equation}
{u_{t + }} = {c_t}\left( {\begin{array}{*{20}{c}}
1\\
0\\
\chi \\
0
\end{array}} \right)\quad \quad {u_{t - }} = {c_t}\left( {\begin{array}{*{20}{c}}
0\\
1\\
0\\
{ - \chi }
\end{array}} \right),\quad 
\label{eqspinorsd2}
\end{equation}
where
\begin{equation}
\chi  = \frac{{{p_t}}}{{{E_t} + {m_t}}} = \frac{{{m_t} - {M_W}}}{{{m_t} + {M_W}}}.
\label{eqchi}
\end{equation}
The right-hand side of eq.\ \eqref{eqchi} can be derived using conservation of linear momentum and energy.  An interaction with both helicities is allowed here because  ${P_L}{u_{t + }} \ne 0$ for  
$v < c,$ as can be seen in figure \ref{figure 3}.  

The current  ${j_2}'$ can be calculated in the 
$O'$   coordinate frame by inserting the spinors of eq.\ \eqref{eqspinorsd} along with the usual form of the Dirac matrices into eq.\ \eqref{eqeps}, which gives
\begin{equation}
{j_2}' = 4\sqrt 2 {c_{{e^ + }}}{c_\nu }{j^ - }.
\label{eqj2p}
\end{equation}
Here  ${j^ - }$ is a unit vector given by
\begin{equation}
{j^ - } = \frac{1}{{\sqrt 2 }}\left( {\begin{array}{*{20}{c}}
0\\
1\\
{ - i}\\
0
\end{array}} \right).\label{eqjminus}
\end{equation}

Since the angle $\theta $  corresponds to a rotation about the y-axis,  ${j_2}'$ can be transformed into the $O$  coordinate frame using
\begin{eqnarray}
{j_2}^ 0 & = & {{j_2}'} ^ 0, \nonumber \\[2pt]
{j_2}^ x  & = & \cos \theta {j_2}'^x - \sin \theta {j_2'^z}  , \nonumber \\[2pt]
{j_2}^y  &=  & {j_2}{'^y}, \nonumber \\[2pt]
{j_2}^z & = & \cos \theta {j_2}{'^z} + \sin \theta {j_2}{'^x},
\label{eq4js}
\end{eqnarray}
where ${j_2}{'^0} = {j_2}{'^z} = 0.$ 

First consider the case where the top quark has negative helicity.  Inserting the spinors ${u_b}$  and ${u_{t - }}$  for the top and bottom quarks into eq.\ \eqref{eqeps} gives a current 
${j_{1 - }}$  of the form
\begin{eqnarray}
{j_{1 - }} & = & 2{c_b}{c_t} \left( {1 + \chi } \right) \left( {j^0} + {j^z} \right) \nonumber
\\[2pt]
 & = & 4{c_b}{c_t}\left( {\frac{{{m_{_t}}}}{{{m_t} + {M_W}}}} \right)\left( {{j^0} + {j^z}} \right).
\label{eqj1minus}
\end{eqnarray}
Here the unit vectors ${j^0}$  and ${j^z}$  are defined by

\begin{equation}
{j^0} \equiv \left( {\begin{array}{*{20}{c}}
1\\
0\\
0\\
0
\end{array}} \right)\quad \quad {j^z} \equiv \left( {\begin{array}{*{20}{c}}
0\\
0\\
0\\
1
\end{array}} \right)\quad .
\label{eqtwojs}
\end{equation}
A similar calculation gives
\begin{eqnarray}
{j_{1 + }} & = &  - 2\sqrt 2 {c_b}{c_t}\left( {1 - \chi } \right){j^ + } \nonumber \\[2pt] & = &  - 4\sqrt 2 {c_b}{c_t}\left( {\frac{{{M_{_W}}}}{{{m_t} + {M_W}}}} \right){j^ +.}
\label{eqj1p}
\end{eqnarray}

Combining equations \eqref{eqMd}, \eqref{eq4js}, and \eqref{eqj1minus} gives the amplitude 
${\mathcal{M}_ - }$  from a top quark with negative helicity as
\begin{equation}
{M_ - } =  - 16\varepsilon {c_{{e^ + }}}{c_\nu }{c_t}{c_b}\left( {\frac{{{m_{_t}}}}{{{m_t} + {M_W}}}} \right)\sin \theta .
\label{eqMminus}
\end{equation}
The corresponding amplitude from a top quark with positive helicity is given by
\begin{equation}
{M_ + } = 16\varepsilon {c_{{e^ + }}}{c_\nu }{c_t}{c_b}\left( {\frac{{{M_{_W}}}}{{{m_t} + {M_W}}}} \right)\left( {1 + \cos \theta } \right).
\label{eqMplus}
\end{equation}
Taking the squares of these amplitudes will give a ${(\sin \theta )^2} = {(\sin \theta ^ *)^2}$  contribution to the angular distribution that is proportional to ${m_t}^2,$  with a smaller 
${(1 + \cos \theta )^2} = {(1 - \cos \theta ^ *)^2}$  contribution that is proportional to ${M_W}^2.$ 

The total probabilities ${p_ - }$  and
${p_ + }$  for these two types of events can be obtained by integrating the squares of  
$\mathcal{M}_ - $  and ${M_ + }$  over a solid angle of  $4\pi ,$ with the result that
\begin{eqnarray}
{p_ - } & = & \frac{1}{2}\alpha {m_t}^2, \nonumber \\[2pt]
{p_ + } & = & \alpha {M_{_W}}^2.
\label{eq2probs}
\end{eqnarray}
Here $\alpha $  is a constant of no interest in what follows.

The angular distribution is conventionally described in terms of the fractional helicities
${f_0}, $  ${f_ -  }, $  and ${f_ + }$  of the W 
 boson, where the longitudinal helicity fraction  ${f_0}$ corresponds to the 
${(\sin \theta ^*)^2}$  contribution to the angular distribution while ${f_ - }$  corresponds to the ${(1 - \cos \theta ^*)^2}$  contribution \cite{aaltonen2010,aaltonen2013,khachatryan2016,kane1992,aguilar2007,czarnecki2010}.  The coefficients in eq.\ \eqref{eq2probs} show that the fractional helicities are given to lowest order by
\begin{eqnarray}
\;{f_0} & = & \frac{r}{{1 + r}}, \nonumber \\[2pt]
{f_ - } & = & \frac{1}{{1 + r}} \nonumber ,\\[2pt]
\;{f_ + } & = & 0,\;\;
\label{eq3fs}
\end{eqnarray}
where the parameter $r$  is defined by
\begin{equation}
r = \frac{1}{2}\frac{{{m_t}^2}}{{{M_{_W}}^2}}.
\label{eqrdef}
\end{equation}

Earlier analyses based on the Standard Model \cite{kane1992,aguilar2007,czarnecki2010} gave the same helicity fractions to lowest order, since the difference between the two theories is proportional to \\
${m_e}^2/{M_W}^2 <  < 1.$  With 
${M_W} = 80.4$  GeV and ${m_t} = 172.8$ GeV, both theories predict  ${f_0} = 0.698,$ ${f_ + } = 0,$  and ${f_{-} } = 0.302$ to lowest order.  The addition of higher-order corrections using the Standard Model gives  ${f_0} = 0.687,$
${f_ + } = 0.0017,$   and ${f_ - } = 0.311$  \cite{czarnecki2010}.  Both sets of theoretical predictions are in good agreement with a fit to the experimental data of figure \ref{figure 8}, which gives  \cite{khachatryan2016}
\begin{eqnarray}
{f_0} & = & 0.705 \pm .013 \pm \nonumber .037, \\[2pt]
{f_ + }& = &  - 0.009 \pm 0.005 \pm 0.021, \nonumber\\[2pt]
{f_ - }  & = & 0.304 \pm 0.009 \pm 0.020.
\label{eq3refsexp}
\end{eqnarray}

These results show that the fact that 
${f_0} > {f_ - } $    is not due to the relativistic enhancement in the longitudinal component of the  
${W^ + }$ boson or the tensor product 
${q_\mu }{q_\nu }/{M_w}^2$  in the propagator, neither of which are present in the alternative quantization approach.  Instead, it is due to the kinematic factor of $\chi $  that appears in the spinor for the top quark.  

If the calculations were performed instead in the ‘t Hooft-Feynman gauge, the contribution from the Goldstone boson would be negligible because, once again, ${m_e}^2/{M_W}^2 <  < 1$   from the vertex factor.  The Goldstone boson equivalence principle states that, in the limit of energies much larger than ${M_W},$ an amplitude involving external longitudinal vector bosons is equivalent to the same amplitude with the external longitudinal vector bosons replaced by the corresponding Goldstone bosons \cite{valencia1990}.  The Goldstone boson equivalence principle does not apply to these calculations, since the W  boson is a virtual state and not an external particle.

The decay of a top quark has previously been analyzed by treating the W  boson as part of the final state, which subsequently decays into the positron and neutrino \cite{peskin2017,kane1992,aguilar2007,czarnecki2010}. The Goldstone boson equivalence principle does apply in that case.  Nevertheless, the coupling of the Goldstone boson to the positron-neutrino final state is proportional to ${m_e}/{M_W}$  and its contribution is negligible once again.

\acknowledgments

The author is grateful to Gordon Baym for discussions regarding conservation of angular momentum in scattering events \cite{baym2024}.

\bibliographystyle{JHEP}
\bibliography{biblio.bib}

\end{document}